\documentclass[conference, letterpaper]{IEEEtran}

\IEEEoverridecommandlockouts

\usepackage{fancyhdr}
\usepackage{xcolor}
\usepackage{epsfig}
\usepackage{threeparttable}
\usepackage{epsf,epsfig}
\usepackage{amsthm}
\usepackage[cmex10]{amsmath}
\usepackage{amssymb, amsfonts}
\usepackage[noadjust]{cite}
\usepackage{dsfont}
\usepackage{subfigure}
\usepackage{color}
\usepackage{soul}
\usepackage{amssymb}
 \usepackage{accents}
  \usepackage{algorithm}
   \usepackage[english]{babel}
    \usepackage{url}
    \usepackage{framed} 
        \usepackage{bm} 

    \usepackage[bottom=0.95in,top=0.7in,left=1.7cm,right=1.7cm]{geometry}

\allowdisplaybreaks[4]

\begin{document}
\newtheorem{theorem}{Theorem}
\newtheorem{acknowledgement}[theorem]{Acknowledgement}
\newtheorem{axiom}[theorem]{Axiom}
\newtheorem{case}[theorem]{Case}
\newtheorem{claim}[theorem]{Claim}
\newtheorem{conclusion}[theorem]{Conclusion}
\newtheorem{condition}[theorem]{Condition}
\newtheorem{conjecture}[theorem]{Conjecture}
\newtheorem{criterion}[theorem]{Criterion}
\newtheorem{definition}{Definition}
\newtheorem{exercise}[theorem]{Exercise}
\newtheorem{lemma}{Lemma}
\newtheorem{corollary}{Corollary}
\newtheorem{notation}[theorem]{Notation}
\newtheorem{problem}[theorem]{Problem}
\newtheorem{proposition}{Proposition}
\newtheorem{scheme}{Scheme}   
\newtheorem{solution}[theorem]{Solution}
\newtheorem{summary}[theorem]{Summary}
\newtheorem{assumption}{Assumption}
\newtheorem{example}{\bf Example}
\newtheorem{remark}{\bf Remark}

\def\qed{$\Box$}
\def\QED{\mbox{\phantom{m}}\nolinebreak\hfill$\,\Box$}
\def\proof{\noindent{\emph{Proof:} }}
\def\poof{\noindent{\emph{Sketch of Proof:} }}
\def
\endproof{\hspace*{\fill}~\qed
\par
\endtrivlist\unskip}
\def\endproof{\hspace*{\fill}~\qed\par\endtrivlist\vskip3pt}

\def\E{\mathsf{E}}
\def\eps{\varepsilon}
\def\phi{\varphi}
\def\Lsp{{\boldsymbol L}}
\def\Bsp{{\boldsymbol B}}
\def\lsp{{\boldsymbol\ell}}
\def\Ltsp{{\Lsp^2}}
\def\Lpsp{{\Lsp^p}}
\def\Linsp{{\Lsp^{\infty}}}
\def\LtR{{\Lsp^2(\Rst)}}
\def\ltZ{{\lsp^2(\Zst)}}
\def\ltsp{{\lsp^2}}
\def\ltZt{{\lsp^2(\Zst^{2})}}
\def\ninN{{n{\in}\Nst}}
\def\oh{{\frac{1}{2}}}
\def\grass{{\cal G}}
\def\ord{{\cal O}}
\def\dist{{d_G}}
\def\conj#1{{\overline#1}}
\def\ntoinf{{n \rightarrow \infty }}
\def\toinf{{\rightarrow \infty }}
\def\tozero{{\rightarrow 0 }}
\def\trace{{\operatorname{Tr}}}
\def\ord{{\cal O}}
\def\UU{{\cal U}}
\def\rank{{\operatorname{rank}}}
\def\acos{{\operatorname{acos}}}

\def\SINR{\mathsf{SINR}}
\def\SNR{\mathsf{SNR}}
\def\SIR{\mathsf{SIR}}
\def\tSIR{\widetilde{\mathsf{SIR}}}
\def\Ei{\mathsf{Ei}}
\def\l{\left}
\def\r{\right}
\def\({\left(}
\def\){\right)}
\def\lb{\left\{}
\def\rb{\right\}}

\setcounter{page}{1}

\newcommand{\eref}[1]{(\ref{#1})}
\newcommand{\fig}[1]{Fig.\ \ref{#1}}

\def\bydef{:=}
\def\ba{{\mathbf{a}}}
\def\bb{{\mathbf{b}}}
\def\bc{{\mathbf{c}}}
\def\bd{{\mathbf{d}}}
\def\bee{{\mathbf{e}}}
\def\bff{{\mathbf{f}}}
\def\bg{{\mathbf{g}}}
\def\bh{{\mathbf{h}}}
\def\bi{{\mathbf{i}}}
\def\bj{{\mathbf{j}}}
\def\bk{{\mathbf{k}}}
\def\bl{{\mathbf{l}}}
\def\bn{{\mathbf{n}}}
\def\bo{{\mathbf{o}}}
\def\bp{{\mathbf{p}}}
\def\bq{{\mathbf{q}}}
\def\br{{\mathbf{r}}}
\def\bs{{\mathbf{s}}}
\def\bt{{\mathbf{t}}}
\def\bu{{\mathbf{u}}}
\def\bv{{\mathbf{v}}}
\def\bw{{\mathbf{w}}}
\def\bx{{\mathbf{x}}}
\def\by{{\mathbf{y}}}
\def\bz{{\mathbf{z}}}
\def\b0{{\mathbf{0}}}

\def\bA{{\mathbf{A}}}
\def\bB{{\mathbf{B}}}
\def\bC{{\mathbf{C}}}
\def\bD{{\mathbf{D}}}
\def\bE{{\mathbf{E}}}
\def\bF{{\mathbf{F}}}
\def\bG{{\mathbf{G}}}
\def\bH{{\mathbf{H}}}
\def\bI{{\mathbf{I}}}
\def\bJ{{\mathbf{J}}}
\def\bK{{\mathbf{K}}}
\def\bL{{\mathbf{L}}}
\def\bM{{\mathbf{M}}}
\def\bN{{\mathbf{N}}}
\def\bO{{\mathbf{O}}}
\def\bP{{\mathbf{P}}}
\def\bQ{{\mathbf{Q}}}
\def\bR{{\mathbf{R}}}
\def\bS{{\mathbf{S}}}
\def\bT{{\mathbf{T}}}
\def\bU{{\mathbf{U}}}
\def\bV{{\mathbf{V}}}
\def\bW{{\mathbf{W}}}
\def\bX{{\mathbf{X}}}
\def\bY{{\mathbf{Y}}}
\def\bZ{{\mathbf{Z}}}

\def\mA{{\mathbb{A}}}
\def\mB{{\mathbb{B}}}
\def\mC{{\mathbb{C}}}
\def\mD{{\mathbb{D}}}
\def\mE{{\mathbb{E}}}
\def\mF{{\mathbb{F}}}
\def\mG{{\mathbb{G}}}
\def\mH{{\mathbb{H}}}
\def\mI{{\mathbb{I}}}
\def\mJ{{\mathbb{J}}}
\def\mK{{\mathbb{K}}}
\def\mL{{\mathbb{L}}}
\def\mM{{\mathbb{M}}}
\def\mN{{\mathbb{N}}}
\def\mO{{\mathbb{O}}}
\def\mP{{\mathbb{P}}}
\def\mQ{{\mathbb{Q}}}
\def\mR{{\mathbb{R}}}
\def\mS{{\mathbb{S}}}
\def\mT{{\mathbb{T}}}
\def\mU{{\mathbb{U}}}
\def\mV{{\mathbb{V}}}
\def\mW{{\mathbb{W}}}
\def\mX{{\mathbb{X}}}
\def\mY{{\mathbb{Y}}}
\def\mZ{{\mathbb{Z}}}

\def\cA{\mathcal{A}}
\def\cB{\mathcal{B}}
\def\cC{\mathcal{C}}
\def\cD{\mathcal{D}}
\def\cE{\mathcal{E}}
\def\cF{\mathcal{F}}
\def\cG{\mathcal{G}}
\def\cH{\mathcal{H}}
\def\cI{\mathcal{I}}
\def\cJ{\mathcal{J}}
\def\cK{\mathcal{K}}
\def\cL{\mathcal{L}}
\def\cM{\mathcal{M}}
\def\cN{\mathcal{N}}
\def\cO{\mathcal{O}}
\def\cP{\mathcal{P}}
\def\cQ{\mathcal{Q}}
\def\cR{\mathcal{R}}
\def\cS{\mathcal{S}}
\def\cT{\mathcal{T}}
\def\cU{\mathcal{U}}
\def\cV{\mathcal{V}}
\def\cW{\mathcal{W}}
\def\cX{\mathcal{X}}
\def\cY{\mathcal{Y}}
\def\cZ{\mathcal{Z}}
\def\cd{\mathcal{d}}
\def\Mt{M_{t}}
\def\Mr{M_{r}}
\def\O{\Omega_{M_{t}}}
\newcommand{\figref}[1]{{Fig.}~\ref{#1}}
\newcommand{\tabref}[1]{{Table}~\ref{#1}}

\newcommand{\var}{\mathsf{var}}
\newcommand{\fb}{\tx{fb}}
\newcommand{\nf}{\tx{nf}}
\newcommand{\BC}{\tx{(bc)}}
\newcommand{\MAC}{\tx{(mac)}}
\newcommand{\Pout}{p_{\mathsf{out}}}
\newcommand{\nnn}{\nn\\}
\newcommand{\FB}{\tx{FB}}
\newcommand{\TX}{\tx{TX}}
\newcommand{\RX}{\tx{RX}}
\renewcommand{\mod}{\tx{mod}}
\newcommand{\m}[1]{\mathbf{#1}}
\newcommand{\td}[1]{\tilde{#1}}
\newcommand{\sbf}[1]{\scriptsize{\textbf{#1}}}
\newcommand{\stxt}[1]{\scriptsize{\textrm{#1}}}
\newcommand{\suml}[2]{\sum\limits_{#1}^{#2}}
\newcommand{\sumlk}{\sum\limits_{k=0}^{K-1}}
\newcommand{\eqhsp}{\hspace{10 pt}}
\newcommand{\tx}[1]{\texttt{#1}}
\newcommand{\Hz}{\ \tx{Hz}}
\newcommand{\sinc}{\tx{sinc}}
\newcommand{\tr}{\mathrm{tr}}
\newcommand{\diag}{\mathrm{diag}}
\newcommand{\MAI}{\tx{MAI}}
\newcommand{\ISI}{\tx{ISI}}
\newcommand{\IBI}{\tx{IBI}}
\newcommand{\CN}{\tx{CN}}
\newcommand{\CP}{\tx{CP}}
\newcommand{\ZP}{\tx{ZP}}
\newcommand{\ZF}{\tx{ZF}}
\newcommand{\SP}{\tx{SP}}
\newcommand{\MMSE}{\tx{MMSE}}
\newcommand{\MINF}{\tx{MINF}}
\newcommand{\RC}{\tx{MP}}
\newcommand{\MBER}{\tx{MBER}}
\newcommand{\MSNR}{\tx{MSNR}}
\newcommand{\MCAP}{\tx{MCAP}}
\newcommand{\vol}{\tx{vol}}
\newcommand{\ah}{\hat{g}}
\newcommand{\tg}{\tilde{g}}
\newcommand{\teta}{\tilde{\eta}}
\newcommand{\heta}{\hat{\eta}}
\newcommand{\uh}{\m{\hat{s}}}
\newcommand{\eh}{\m{\hat{\eta}}}
\newcommand{\hv}{\m{h}}
\newcommand{\hh}{\m{\hat{h}}}
\newcommand{\Po}{P_{\mathrm{out}}}
\newcommand{\Poh}{\hat{P}_{\mathrm{out}}}
\newcommand{\Ph}{\hat{\gamma}}
\newcommand{\mat}[1]{\begin{matrix}#1\end{matrix}}
\newcommand{\ud}{^{\dagger}}
\newcommand{\C}{\mathcal{C}}
\newcommand{\nn}{\nonumber}
\newcommand{\nInf}{U\rightarrow \infty}

\title{\huge Bayesian Over-the-Air FedAvg via Channel Driven \\ Stochastic Gradient Langevin Dynamics}

\author{ \IEEEauthorblockN{Boning Zhang\IEEEauthorrefmark{1}, Dongzhu Liu\IEEEauthorrefmark{1}, Osvaldo Simeone\IEEEauthorrefmark{2}, and Guangxu Zhu\IEEEauthorrefmark{3} 
\thanks{The work of O. Simeone was supported by
the European Research Council (ERC) through European Union’s Horizon 2020
Research and Innovation Programme under Grant 725731, by  an Open Fellowship of the EPSRC with reference
EP/W024101/1, and by the European Union’s Horizon Europe project
CENTRIC (101096379). }
}
\IEEEauthorblockA{\IEEEauthorrefmark{1}{School of Computing Science, University of Glasgow} \\
\IEEEauthorrefmark{2}{ Department of Engineering, King's College London} \\
\IEEEauthorrefmark{3}{Shenzhen Research Institute of Big Data} \\
Email: b.zhang.6@research.gla.ac.uk,
    dongzhu.liu@glasgow.ac.uk, 
    osvaldo.simeone@kcl.ac.uk,
    gxzhu@sribd.cn}
}

\maketitle

\begin{abstract}
The recent development of scalable Bayesian inference methods  has renewed interest in the adoption of Bayesian learning as an alternative to conventional frequentist learning that offers improved model calibration via uncertainty quantification. Recently, federated averaging Langevin dynamics (FALD) was introduced as a  variant of federated averaging that can efficiently implement distributed Bayesian learning in the presence of noiseless communications. In this paper, we propose wireless FALD (WFALD), a novel protocol that realizes FALD in wireless systems by integrating over-the-air computation and channel-driven sampling for Monte Carlo updates.  Unlike prior work on wireless Bayesian learning, WFALD enables (\emph{i}) multiple local updates between communication rounds; and (\emph{ii}) stochastic gradients computed by mini-batch. A convergence analysis is presented in terms of the 2-Wasserstein distance between the samples produced by WFALD and the targeted global posterior distribution. Analysis and experiments show that, when the signal-to-noise ratio is sufficiently large,  channel noise can be fully repurposed for Monte Carlo sampling, thus entailing no loss in performance. 
\end{abstract}


\section{Introduction}

With the increasingly widespread use of machine learning tools in sensitive applications, the reliability of deep learning techniques has come under intense scrutiny, revitalizing research on uncertainty quantification and calibration (see, e.g., \cite{guo2017calibration,abdar2021review}). The golden standard of well-calibrated machine learning techniques is set by  \emph{Bayesian learning}, which treats model parameters are random variables. Recent developments in \emph{scalable} Bayesian inference have made Bayesian learning techniques practical contenders for applications that require uncertainty quantification \cite{angelino2016patterns,simeone2022machine,wilson2020bayesian,khan2021bayesian}.  This paper focuses on the efficient distributed implementation of Bayesian learning in wireless systems (see Fig. 1).

\emph{Federated learning} (FL) refers to distributed learning protocols that enables multiple devices to collaboratively train a statistical model without directly sharing the local data sets \cite{mcmahan2017communication, konevcny2016federated}. The training process involves communicating model parameters between the server and devices. To alleviate the communication cost, a popular approach, popularized by \emph{federated averaging} (FedAvg) \cite{mcmahan2017communication},  is to reduce the communication frequency by allowing  devices perform multiple local updates before sending the local model to the central server  for global aggregation. Implementing FL over wireless channels provides another opportunity to enhance communication efficiency via \emph{over-the-air computing} (AirComp) \cite{zhu2019broadband, liu2020privacy,cao2022transmission}. AirComp leverages  uncoded transmission of the local updates over a non-orthogonal multiple access (NOMA) channel, using the superposition property of wireless channels to enable the efficient model aggregation on air.

\begin{figure}[t]
\centering
\includegraphics[width=8cm]{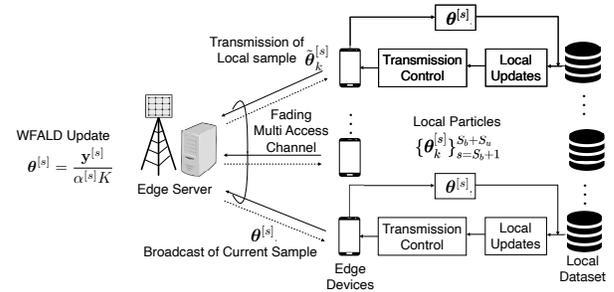}
\vspace{-2mm}
\caption{The proposed Wireless federated averaging Langevin dynamics (WFALD) protocol implements distributed Bayesian learning via over-the-air channel-driven sampling and multiple local updates between global aggregation steps. }
\vspace{-8mm}
\label{Fig: FL system}
\end{figure}

Most existing FedAvg implements \emph{frequentist learning},  where the target is to estimate an optimal model that minimizes the empirical loss by stochastic gradient descent (SGD) algorithms \cite{sery2021over}. One of the simplest ways to implement Bayesian learning is through  a variant of SGD known as \emph{stochastic gradient Langevin dynamics} (SGLD) \cite{welling2011bayesian}. By \emph{injecting Gaussian noise} into the SGD update, the model  parameters produced by SGLD can approximate distribution of the Bayesian posterior density. 


Federated protocols that implement Bayesian learning are less well studied, and are currently subject to intense research \cite{cao2023bayesian}. Distributed SGLD was first introduced in \cite{ahn2014distributed} based on gradient averaging, allowing for a single local iteration between global aggregation rounds. Multiple local updates and model parameter averaging were allowed in the implementations proposed in  \cite{deng2021convergence} and  \cite{plassier2022federated}. In particular, the state-of-the-art approach in \cite{plassier2022federated}, referred to as \emph{federated averaging Langevin dynamics} (FALD), was proved to have desirable convergence properties. Another research direction aims at reducing the communication cost of federated SGLD by quantizing the update gradients \cite{vono2021qlsd}. 

The prior work discussed above on distributed SGLD algorithms assumes noiseless communications. References  \cite{liu2021channel, WFLMC, zhang2022leveraging} introduced the idea of \emph{channel-driven sampling}, whereby channel noise is repurposed as a resource for over-the-air Bayesian learning via Monte Carlo sampling. Specifically, reference  \cite{WFLMC} implemented a basic version of decentralized Langevin dynamics that allows for a single local update round between global aggregation rounds and for full-batch gradient updates. A fully decentralized version of the protocol was also studied in \cite{barbieri2022channel}. 

In this context, the main contributions of this work are as follows.
\begin{itemize}
    \item[$\bullet$] \textbf{Wireless federated averaging Langevin dynamics (WFALD):} We introduce wireless federated averaging Langevin dynamic (WFALD), a novel Bayesian learning algorithm that integrates over-the-air computing with channel-driven sampling, while enabling (\emph{i}) multiple local updates between communication rounds; and (\emph{ii}) the use of stochastic gradients. WFLAD requires minor modifications to the standard over-the-air implementation of FedAvg.
    \item[$\bullet$] \textbf{Convergence analysis:} We analyze the convergence of WFALD in terms of  2-Wasserstein distance. The analytical bound reveals that increasing global aggregation frequency in the low-signal-to-noise ratio (SNR) regime may not always be helpful, since excessive  channel noise cannot be fully repurposed for MC sampling. 
    \item[$\bullet$] \textbf{Experiments:} We demonstrate the performance of WFLAD across different global aggregation frequencies and SNR settings to verify the analytical results. 
\end{itemize}

\section{System Model}

As shown in Fig. \ref{Fig: FL system}, we consider a distributed learning system comprising a single-antenna edge server and $K$ edge devices. Each device $k$ has its own local dataset $\cD_k$ encompassing $N_k$  data samples $\cD_k=\{ \bd_{k,n} \}_{n=1}^{N_k}$. Accordingly, we denote global dataset as $\cD=\bigcup_{k=1}^K \cD_k= \{\bd_{n}\}_{n=1}^{N}$ with $N=\sum_{k=1}^K N_k$. As we will detail in this section, the single-antenna  devices communicate to the server via non-orthogonal multi-access (NOMA) channel with uncoded modulation, as in \cite{WFLMC}. The goal of the system is to carry out gradient-based Monte Carlo (MC) sampling via a FedAvg paradigm that leverages channel-driven sampling \cite{WFLMC, liu2021channel} and over-the-air computing \cite{zhu2020over}. As explained in Sec. I, unlike \cite{WFLMC}, which focused on single gradient descent (GD) update between global aggregation steps, we allow for multiple local stochastic gradient descent (SGD)  updates.

\subsection{Learning Problem} \label{sec: learning protocol}
We consider the learning model defined by a likelihood function  $p( \bd |{\bm \theta}  )$ and a prior distribution $p(\bm \theta)$.  Accordingly, the likelihood at device $k$  is given as follows 
\vspace{-1mm}
\begin{align}
 p(\cD_k|{\bm \theta}) = \prod_{n=1}^{N_k}  p(\bd_{n,k} |{\bm \theta}).
\end{align}

The goal of the learning system is to estimate averages of quantities of interest, such as the ensemble predictive distribution (see, e.g., \cite[Chapter 12]{simeone2022machine}), with respect to the \emph{global posterior distribution} 
\vspace{-2mm}
\begin{align}
 p({\bm \theta}|\cD)  \propto p({\bm \theta}) \prod_{k=1}^K  p(\cD_k | {\bm \theta}).  \label{eq: posterior} \vspace{-2mm}
 \end{align} 
 The global posterior can be decomposed as a product of $K$ local  sub-posteriors as $ p({\bm \theta}|\cD)  \propto \prod_{k=1}^K  \tilde{p}({\bm \theta}|\cD_k)$, with each \emph{sub-posterior} given by \vspace{-2mm}
\begin{align}
 &\tilde{p}({\bm \theta}|\cD_k)  \propto p({\bm \theta})^{1/K}  p(\cD_k | {\bm \theta}). \label{eq: local sub-posterior} \vspace{-2mm}
\end{align} 
The global posterior (\ref{eq: posterior}) is related to the conventional local regularized training loss  \vspace{-2mm}
\begin{align}\label{eq: local func}
f_k(\bm \theta)=-\log p(\cD_k|{\bm \theta}) -\frac{1}{K}\log p(\bm \theta)
\end{align} for all devices $k$.  In fact, we have the proportionality relation $p({\bm \theta}|\cD)\propto \exp(-f(\bm \theta))$, where the global cost function  \vspace{-2mm}
\begin{align}\label{eq: gb func}
f(\bm \theta)= \sum_{k=1}^K f_k(\bm \theta)
\end{align}  corresponds to the standard \emph{global training loss} addressed by frequentist federated learning.

\subsection{Stochastic Gradient Langevin Dynamics}
\emph{Stochastic gradient Langevin dynamics} (SGLD) is a stochastic gradient-based MCMC sampling scheme. Like all MCMC techniques, the goal is to produce a number of \emph{particles} $\bm \theta^{[s]}$ with $s=1,...,S$ that are approximately distributed according to a posterior distribution of interest, here, the global posterior (\ref{eq: posterior}). Samples  $\{\bm \theta^{[s]}\}_{s=1}^S$ can then be used to obtain estimates of averages with respect to the posterior $ p({\bm \theta}|\cD)$ of some function $g(\bm \theta)$ as $1/S \sum_{s=1}^S g(\bm \theta^{[s]})$. As a notable example, with  $g(\bm \theta)= p(\bm d| \bm \theta)$, this average yields the ensemble Bayesian predictor regarding the test data $\bm d$. 

Starting with an arbitrary particle $\bm \theta^{[0]}$, SGLD evaluates a stochastic estimate of the gradient using a mini-batch $\cC^{[s]} \subseteq \cD$ of size $p_b N$ from the global data set for some $p_b\in(0,1]$ as 
\vspace{-1mm}
\begin{align} \label{eq:stochastic gradient}
\hat{\nabla} f(\boldsymbol{\theta}^{[s]}) \! = \! - \frac{1} {p_b} \!\!\! \sum_{\bd_n \in \cC^{[s]}} \!\!\!\! \nabla \log p(\mathbf{d}_{n} \! \! \mid \boldsymbol{\theta}^{[s]}) \! - \! \frac{1}{K} \nabla \log p(\boldsymbol{\theta}^{[s]}).  \vspace{-4mm}
\end{align} 
\vspace{-1mm}
Then, it applies  a noise-perturbed  SGD update
 \begin{align} \label{eq: LMC}
\text{(SGLD)} \quad {\bm \theta}^{[s+1]}={\bm \theta}^{[s]} -\eta  \hat{\nabla} f({\bm \theta}^{[s]})  + \sqrt{2\eta} {\bm \xi}^{[s]},  
\end{align}  
where  ${\bm \xi}^{[s]}$ for $s=1,2,...$ is an i.i.d. sequence of standard Gaussian variables $\cN(0, \bI)$. The SGLD update (\ref{eq: LMC}) requires access to the entire data set $\mathcal{D}$, and hence it cannot be directly implemented in a decentralized setting.

\subsection{Federated Averaging Langevin Dynamics (FALD)}\label{sec:FALD}
Reference \cite{deng2021convergence} recently introduced a federated version of SGLD -- referred to as FALD -- that incorporates multiple local updates between global FedAvg-type aggregation steps. Devices operate according to a common time step $s=1,2,...$, and are assumed to have a source of common randomness, i.e., a common seed, to generate shared random variables.  

At each iteration $s$,  device $k$ randomly draws $p_b N_k$  data samples, for some $p_b \in (0, 1]$, as a batch $\cC_{k}^{[s]} \subseteq   \cD_k$ from its local dataset, and computes stochastic gradient as in \eqref{eq:stochastic gradient}. Then, it applies the SGLD-like update 
\begin{align}\label{eq: inter local particle}
\tilde{\bm{\theta}}_{k}^{[s+1]} =\bm{\theta}_k^{[s]}-\eta \hat{\nabla} f_k(\boldsymbol{\theta}_{k}^{[s]}) +  \sqrt{2\eta} {\bm \xi}_k^{[s]}, 
\end{align}  
where  the i.i.d. additive noise $\bm \xi_k^{[s]}$ with $s=1,2,...$ is generally correlated across devices $k$. Specifically, the correlation among additive noise terms $\{\bm \xi_k^{[s]}\}_{k=1}^{K}$ is controlled by a parameter $\tau^{[s]}\in [0,1]$ as  \vspace{-2mm}
\begin{align} \label{eq: noise decomp}
{\bm \xi}_k^{[s]} = \sqrt{\frac{\tau^{[s]}}{K}} \tilde{ {\bm \xi}}_c^{[s]}   + \sqrt{1-\tau^{[s]}}    \tilde{ {\bm \xi}}_k^{[s]},  
\end{align} where noise term $\tilde{ {\bm \xi}}_c^{[s]}  \sim \cN(0, \bI)$ is shared across all devices, while  $\tilde{ {\bm \xi}}_k^{[s]} \sim \cN(0, \bI)$ is independent across different device $k$.

After each local update, at each iteration $s$, all devices collectively decide, using common randomness, whether to communicate or not with the server. Accordingly, a common Bernoulli random variable $B^{[s]}$ with probability $p_c$ is drawn at all devices. When $B^{[s]}=1$, each device $k$ communicates its current updated particle $\tilde{\bm{\theta}}_k^{[s]}$ in (\ref{eq: inter local particle}) to the server; while when $B^{[s]}=0$ no communication takes place, and all devices move on to the next iteration $s+1$. After receiving particles from all devices, the server obtains a global sample by applying the standard FedAvg update \vspace{-2mm}
 \begin{align}\label{eq: global aggregation}
\text{(Global Aggregation)} \quad {\bm \theta}^{[s+1]} = \frac{1}{K}\sum_{k=1} ^K  \tilde{\bm{\theta}}_k^{[s+1]}
\end{align}  
and broadcasts it to all devices to be used as the initialization for the next iteration. 

Specifically, the local particle ${\bm{\theta}}_k^{[s+1]}$ to be used as the initialization for the next iteration is set as  \vspace{-1mm}
\begin{align} \label{eq: local update}
\text{(Update)} \ \bm{\theta}_k^{[s+1]} = (1 - B^{[s]})\tilde{\bm{\theta}}_{k}^{[s+1]} +  \frac{B^{[s]}}{K} \sum_{k=1}^{K} \tilde{\bm{\theta}}_k^{[s+1]}. 
\end{align} 
 Note that by  \eqref{eq: noise decomp}, after global aggregation \eqref{eq: global aggregation}, the system implements the SGLD update (\ref{eq: LMC}) with effective learning rate $\eta/K$. 
 

\subsection{Communication Model}

In the uplink,  devices communicate to the edge server on the shared non-orthogonal multiple access (NOMA) channel. Each device implements \emph{uncoded analog transmission} to realize efficient particle aggregation step (\ref{eq: global aggregation}) via over-the-air computing \cite{zhu2019broadband, liu2020privacy,cao2022transmission}, as well as to exploit the channel noise as the correlated noise term in (\ref{eq: noise decomp}). As we detail in the next section, we emphasize that the latter idea is a key novel contribution to this work. As in prior art \cite{liu2020privacy, cao2022transmission}, we assume perfect channel state information (CSI) at all nodes, which enables power control for over-the-air computing. 

We assume a block flat-fading channel where the channel coefficients remain constant for $d$ channel uses. This assumption enables the transmission of $d$-dimensional sample vectors $\{\bm{\theta}_k^{[s]}\}_{k=1}^K$ within a block. Symbol-level synchronization among all devices can be achieved by using standard protocols such as the timing advance procedure in LTE and 5G NR \cite{mahmood2019time}. In the $s$-th communication round, the   signal received at the server is  \vspace{-1mm}
\begin{equation}\label{eq: channel}
\mathbf{y}^{[s]}=\sum_{k=1}^K h_k^{[s]} \mathbf{x}_k^{[s]}+\bz^{[s]} \vspace{-3mm}
\end{equation}
where $h_k^{[s]}$ is the channel gain for device $k$ in round $s$, $\mathbf{x}_k^{[s]} \in \mathbb{R}^d$ is the transmit signal containing the information of the local particles, and $\bz^{[s]}$ is channel noise i.i.d. according to distribution $\mathcal{N}\left(0, N_0 \mathbf{I}\right)$. Each device has a transmit power constraint 
\begin{equation}\label{eq: power constraint}
\|\mathbf{x}_k^{[s]}\|^2 \leq P, \ \text{for all $k$}
\end{equation}
accounting for each communication block. We define signal-to-noise ratio as $\SNR=P/(d N_0)$. 
As in most other related papers, we assume noiseless downlink communication. This assumption is practically well justified when the edge server communicates through a base station with less stringent power and bandwidth constraints than the devices. 

\subsection{Assumptions on the Local Cost Function}
Finally, we list several standard assumptions  (see, e.g., \cite{WFLMC}) we make on the local cost function $f_k(\bm \theta)$ in \eqref{eq: local func} and on its gradient.
\vspace{-1mm}
\begin{assumption}[Smoothness]\label{assumption: Lipschitz}\emph{
The local cost function $f_k(\bm \theta)$, $k=1,..., K$,  is smooth with constant $L>0$, that is, it is continuously differentiable and the gradient $\nabla f_k(\bm \theta)$  is Lipschitz continuous with constant $L$, i.e.,   
\begin{align}\label{eq: Lipschitz} 
\|\nabla f_k({\bm \theta})-\nabla f_k({\bm \theta}')\| \leq L \| {\bm \theta}-{\bm \theta}'\|,  \quad \forall {\bm \theta},{\bm \theta}' \in \mathbb{R}^d.  \nn
\end{align} 
}
\end{assumption}
\vspace{-2mm}
\begin{assumption}[Strong Convexity]\label{assumption: Strong Convexity}\emph{The following inequality holds for any local cost function $f_k(\bm \theta)$, $k=1,\cdots, K$, with some constant $\mu>0$  
 \begin{equation}\label{eq: PL ineq}
\l[\nabla f_k(\bm \theta)\!- \! \nabla f_k({\bm \theta}')\r]^{\sf T} \!\! ({\bm \theta}-{\bm \theta}') \! \geq \! \mu \|{\bm \theta} - {\bm \theta}'\|^2,  \forall {\bm \theta},{\bm \theta}' \in \mathbb{R}^d.  
 \end{equation}
 }
\end{assumption}      

\vspace{-2mm}
\begin{assumption}[Unbiased and Variance Bounded Stochastic Gradient] \label{A2} 
\emph{The stochastic gradient $\hat{\nabla} f_k(\bm{\theta})$ is an unbiased estimate of the local gradient ${\nabla} f_k(\bm{\theta})$, and the variance is bounded by $\sigma_k^2$ as  \vspace{-1mm}
\begin{align}
\mathbb{E}\! \left[\left\| \hat{\nabla} f_k(\bm{\theta})\! - \! \nabla f_k (\bm{\theta}) \right\|^2 \right] \! \leq \! \sigma_k^2,   \forall {\bm \theta}\in \mathbb{R}^d, k = 1,\! \cdots\!,\! K.  
\end{align}
}
\end{assumption}

\begin{assumption}[Bounded Local Gradient]  \label{A3} 
\emph{The local gradient is bounded as   
\begin{align}
\l\|\nabla f_k({\bm \theta}) \r \| \leq G ,   \forall {\bm \theta}\in \mathbb{R}^d, \ k=1,\cdots, K.   \end{align}
}
\end{assumption}
\vspace{1mm}
\section{Wireless Federated Averaging Langevin Dynamics}
In this section, we introduce the proposed WFALD protocol, which integrates FALD \cite{deng2021convergence}, reviewed in the previous section, with channel-driven sampling \cite{WFLMC, liu2021channel} and over-the-air computing \cite{zhu2020over}. The goal is to define an efficient  distributed Bayesian learning protocol that requires only  minor changes to the conventional frequentist federated learning based on over-the-air computing. Accordingly, unlike \cite{WFLMC}, we allow for multiple local updates as dictated by the random scheduling mechanism described in Sec. \ref{sec:FALD}. The main idea of this work is to leverage channel noise during global aggregation steps as the common noise term in the FALD update (\ref{eq: noise decomp}).

\subsection{Overview of WFALD}\label{sec: WFALD}

As proved in \cite[Theorem 1]{plassier2022federated}, the addition of  the common noise term in \eqref{eq: noise decomp} is critical to reducing the discrepancy between the  distribution of the aggregated particles in (\ref{eq: global aggregation}) and the target global posterior (\ref{eq: posterior}). In the implementation proposed in \cite{deng2021convergence}, this requires nodes to share enough common randomness to be able to generate i.i.d. Gaussian random variables at each iteration $s$ as in \eqref{eq: noise decomp}. WFALD is based on the following observation. 


At any iteration $s$ with global aggregation, i.e., with $B^{[s]}=1$, each device $k$ communicates the updated local particle
 \begin{equation}\label{eq: tx signal}
\mathbf{x}_{k}^{[s]} = \alpha_{k}^{[s]} \big[\bm{\theta}_k^{[s]}-\eta \hat{\nabla} f_k(\boldsymbol{\theta}_{k}^{[s]}) \big]
 \end{equation}
 using uncoded transmission on the NOMA channel (\ref{eq: channel}) with some power control parameter $\alpha_{k}^{[s]}$. The design of power control, as detailed in the following subsection, ensures that the received signal \eqref{eq: channel}  can be scaled by a factor $c^{[s]}$ to approximate the updated particle $\bm{\theta}^{[s+1]}$ in \eqref{eq: global aggregation} with $\tau^{[s]}=1$, i.e., \vspace{-4mm}
\begin{align}\label{eq: WFALD comm}
\frac{\by^{[s]}}{c^{[s]}} \approx    \frac{1}{K}\sum_{k=1}^K\bm{\theta}_k^{[s]}-\frac{\eta}{K} \sum_{k=1}^K \hat{\nabla} f_k(\boldsymbol{\theta}_{k}^{[s]})  + \sqrt{\frac{2\eta}{K}}\tilde{\bm \xi}_c.  
\end{align}  

  For iterations $s$ when no global aggregation is done, i.e., $B^{[s]}=0$, WFALD stipulates that each device $k$ apply the SGLD update \eqref{eq: inter local particle} with $\tau^{[s]}=0$, i.e.,  adding only the independent local noise terms $\tilde{\bm \xi}_k^{[s]}$. This choice is made for simplicity of implementation, and it may be removed if the devices have sufficient common randomness to generate shared Gaussian noise.  
  
   The outlined design of WFALD has the double advantage of reducing the requirement of shared global randomness, and, even more importantly, of mitigating the impact of channel noise, which is repurposed as a useful resource for learning \cite{WFLMC, liu2021channel}.     Any  additional noise is treated as a nuisance, requiring a modification of the analysis of convergence presented in \cite{plassier2022federated}, as we discuss next.

\subsection{Signal Design and Power Control}
As anticipated in the previous subsection, for iterations with $B^{[s]}=1$, all devices transmit their updated local particles per \eqref{eq: tx signal}. 
As in \cite{liu2020privacy}, we implement channel inversion and select the power control parameter as $\alpha_k^{[s]}=\alpha^{[s]}/ h_k^{[s]} $ for signal alignment,  where $\alpha^{[s]}>0$ is a common gain parameter to be designed. In order to estimate the global update \eqref{eq: global aggregation}, the received signal \eqref{eq: channel} is then scaled as in  \eqref{eq: WFALD comm} with 
 \begin{equation}
 c^{[s]}=K \alpha^{[s]}
	\label{scale}
\end{equation} 


To explain the rationale behind the design \eqref{scale}, we plug \eqref{eq: channel}, \eqref{eq: tx signal}, and \eqref{scale} into \eqref{eq: WFALD comm}, yielding
\vspace{-1mm}
\begin{align}
 &\text{(Wireless Global Aggregation)} \nn\\
  &\bm{\theta}^{[s+1]} =\frac{1}{K} \sum_{k=1}^K \bm{\theta}_k^{[s]} -  \frac{\eta}{K} \sum_{k=1}^K \hat{\nabla} f_k(\boldsymbol{\theta}_{k}^{[s]}) + \frac{\bz^{[s]}}{\alpha^{[s]}K}  \\
& = \frac{1}{K}\! \sum_{k=1}^K \bm{\theta}_k^{[s]}\!\! - \!\! \frac{\eta}{K}\! \sum_{k=1}^K \! \hat{\nabla} f_k(\boldsymbol{\theta}_{k}^{[s]})\! + \! \sqrt{\frac{2 \eta}{ K }} \tilde{\bm \xi}_c^{[s]} \! + \! \sqrt{\beta^{[s]}} \! \bm{\Delta}^{[s]}  \label{eq: update residual nosie}
\end{align}
where   $\bm{\Delta}^{[s]} \sim \mathcal{N}\left(0, \mathbf{I}_d\right)$ are i.i.d. over the iteration index $s$, and we have  defined 
\vspace{-2mm}
\begin{equation}
	\beta^{[s]} = \max\left\{0, \frac{N_{0}}{(\alpha^{[s]} K)^2} - \frac{2\eta}{K}\right\}. 
\end{equation}
Comparing \eqref{eq: update residual nosie} with \eqref{eq: WFALD comm} reveals that the estimate \eqref{scale} corresponds to the global update \eqref{eq: WFALD comm} apart from the addition of the noise term $\sqrt{\beta^{[s]}}\bm{\Delta}^{[s]}$. 

The additional noise term can be forced to be equal zero by setting $\alpha^{[s]}=\sqrt{N_0/(2\eta K)}$. However, this choice may not be always feasible due to the  power constraint \eqref{eq: power constraint}, which yields the condition  $\alpha^{[s]} \leq \min_k   {\sqrt{P}\|h_k^{[s]}\|}/{ \| \bm{\theta}_k^{[s]}-\eta \hat{\nabla} f_k(\boldsymbol{\theta}_{k}^{[s]})\|}  $. Taking these considerations into account, WFALD selects the  power control gain as \vspace{-1mm}
\begin{align}\label{eq: power gain setting}
\alpha^{[s]}= \min\l\{\sqrt{\frac{N_0}{2\eta K}}, \min_k    \frac{\sqrt{P}\|h_k^{[s]}\|}{ \| \bm{\theta}_k^{[s]}-\eta \hat{\nabla} f_k(\boldsymbol{\theta}_{k}^{[s]})  \|}   \r\}. 
\end{align}

\subsection{Convergence Analysis}
To analyze the quality of the samples $ \{\bm{\theta}_k^{[s]}\}_{k=1}^K$  produced by  WFALD after $s$ iterations, as in \cite{plassier2022federated}, we study the distribution $p(\tilde{\bm \theta}^{[s]})$  of the averaged local particles \vspace{-2mm}
\begin{align}\label{analysis: average local particles}
\tilde{\bm{\theta}}^{[s]} &= \frac{1}{K} \sum_{k=1}^{K} \bm{\theta}_k^{[s]}.  
\end{align} 
Specifically, we analyze the standard 2-Wasserstein distance $W_2(p(\tilde{\bm \theta}^{[s]}),p({\bm\theta}|\cD))$ between distribution $p(\tilde{\bm \theta}^{[s]})$  and the target global posterior $p({\bm\theta}|\cD)$ in \eqref{eq: posterior}, on which we provide an upper bound in the following theorem. 


\begin{theorem}\label{theorem1}
\emph{For a learning rate  $0 < \eta \leq 2/L$, under Assumptions \ref{assumption: Lipschitz}-\ref{A3}, after $s$ iterations, the 2-Wasserstein distance between the distribution of the averaged local samples produced by WFALD and the global posterior is upper bounded as
\vspace{-2mm}
	\begin{equation}\label{eq: Theorem1}
		\begin{aligned}
			& \mathbf{W}_2^2(p(\tilde{\bm \theta}^{[s]}), p(\bm{\theta}|\mathcal{D}))  \leq \left(\frac{1 + \gamma}{2}\right)^{2s} \mathbf{W}_2^2(p(\tilde{\bm \theta}^{[0]}), p(\bm{\theta}|\mathcal{D}) ) + \\
             & \sum_{j=0}^{s - 1} \left(\frac{1 + \gamma} {2}\right)^{2(s-j)}\!\! p_c \beta^{[s]} d + \frac{8(1 + \gamma)}{3(1- \gamma)^2} \left[ \frac{\eta^4 L^3 d}{3K} + \eta^3 L^2 d + \right. \\
           & \left. \Big(\frac{\eta^2}{K} \! + \!\frac{4\eta^4 L^2}{Kp_c} \Big)\!\! \sum_{k=1}^{K} \! \sigma_{k}^2 \! + \! \frac{6\eta^4L^2 G^2}{p_c^2} \! + \! \frac{4 \eta^3 L^2(K \!- \! 1)  d}{Kp_c} \right]\! , 
		\end{aligned}
	\end{equation}
where $\gamma=1-\eta \mu$ for $0 < \eta \leq 2 /(\mu + L)$ and $\gamma = \eta L-1$ for $2 /(\mu + L) \leq \eta \leq 2 / L$; $\sigma_k^2$ is the bounded variance of stochastic local gradients. 
}
\end{theorem}



As detailed in the appendix, this result follows by modifying the proof of \cite[Proposition 1]{WFLMC} by accounting for the divergence incurred by multiple stochastic local updates, also known as the client drift \cite{karimireddy2020scaffold}. 

The main insights from the bound (\ref{eq: Theorem1}) concern the role of the global aggregation rate $p_c$. In particular, in the high-SNR regime, where $\beta^{[s]}=0$, setting $p_c=1$ is seen to reduce the convergence bound, indicating that communication is always useful. In contrast, when the SNR is below a threshold,  larger values of  $p_c$  may increase the bound, suggesting that  excessively frequent communication may be harmful to the convergence. This is due to the presence of the non-zero residual channel noise with power $\beta^{[s]}$ in the update \eqref{eq: update residual nosie}. Accordingly, in the low-SNR regime, there exists an optimal value of the communication frequency $p_c$ that best trades off the client drift error and the residual channel noise. 

We finally note that the result in  \cite[Proposition 1]{WFLMC} pertains a simpler protocol that applies global aggregation at every step, i.e., $p_c=1$, and with full-batch gradients, i.e., $p_b=1$ which implies $\sigma_k=0$. Therefore, the bound \eqref{eq: Theorem1} does not reduce to that in \cite[Proposition 1]{WFLMC}, since, as explained in Sec. \ref{sec: WFALD}, the two schemes apply a different transmission strategy.

\section{Numerical Results}


To bring some quantitative insights into the performance of WFALD, as  in \cite{WFLMC}, we consider a standard Gaussian linear regression problem with the likelihood
\vspace{-1mm}
\begin{equation} \label{equ:linear_model}
p\left(v_n \mid \boldsymbol{\theta}, \mathbf{u}_n\right)=\frac{1}{\sqrt{2 \pi}} e^{-\frac{1}{2}\left(v_n-\boldsymbol{\theta}^{\top} \mathbf{u}_n\right)^2},
\end{equation}
and prior $p({\bm \theta})$ assumed to follow Gaussian distribution $\mathcal{N}(0, \mathbf{I})$.  Accordingly, the posterior $p({\bm \theta}|\cD)$ is Gaussian distribution  $\cN\big((\bU\bU^{\sf T}+\bI)^{-1}\bU\bv,(\bU\bU^{\sf T}+\bI)^{-1}\big)$,  where $\bU=[\bu_1,\cdots,\bu_N]$ is the data matrix and $\bv\!\!=\!\![v_1, \cdots, v_N]^{\sf T}$ is the vector of target variables.  We use a synthetic dataset $\left\{\mathbf{d}_n=\left(\mathbf{u}_n, v_n\right)\right\}_{n=1}^N$ with $N = 1200$ following the model (\ref{equ:linear_model}) with covariates $\mathbf{u}_n \in \mathbb{R}^{d}$ drawn i.i.d. from Gaussian $\mathcal{N}\left(0, \mathbf{I}\right)$ where $d = 5$. The ground-truth model parameter generating the  target variables  $v$ through the likelihood \eqref{equ:linear_model}  is ${\bm \theta}^{\star}$ = [$-$0.0615, $-$1.6057, 1.7629, 1.0240, $-$1.5902]$^{\sf T}$. Unless stated otherwise, the dataset is evenly distributed to $K=30$ devices; the channel $h_k^{[s]}$ is set to 1 for all communication rounds and devices; the learning rate is $\eta = 3 \times 10^{-3}$; and $p_b$ is 0.4. 







We run  $S=200$ iterations. In order to evaluate the averages of interest over the particles, we assume a burn-in period of $S_{b} = 100$, hence only retaining the remaining  $S_u = S - S_b = 100$ samples for evaluation. As a test function, we consider two choices. First, we consider the identity $g(\bm \theta)=\bm \theta$, and thus we compare the average $1/S_u \sum_{s=S_b + 1}^{S_b + S_u} {\bm \theta}_k^{[s]}$ of the particles generated by WFALD at each device $k$ after the burn-in period with the true average ${\bm \mu}_p= (\bU\bU^{\sf T}+\bI)^{-1}\bU\bv $ of the global posterior (\ref{eq: posterior}). To this end, we measure mean squared error between empirical and true averages over all devices as 
\vspace{-1mm}
\begin{align}\label{eq:MSE}
   \text{MSE} = \frac{1}{K} \sum_{k=1}^{K} \Big\| \frac{1}{S_u} \sum_{s=S_b + 1}^{S_b + S_u} {\bm \theta}_k^{[s]} - {\bm \mu}_p \Big\|^2 .
\end{align} 
Then, in order to enable a comparison with conventional frequentist wireless FedAvg (WFedAvg), we consider the test function  $g(\bm \theta)=p(v|\bm{\theta},\mathbf{u})$ and we evaluate the performance in terms of average quadratic error on a test set of 500 examples per device when using the expectation produced by the ensemble average predictor $\frac{1}{S_u} \sum_{s=S_b + 1}^{S_b + S_u} p(v|{\bm \theta}_k^{[s]},\mathbf{u}) $ for Bayesian learning, and by  the predictor $p(v|\bm \theta_k^{[S]},\mathbf{u})$ obtained with the last iterate for WFedAvg. WFedAvg is implemented using the same mini-batch size and frequency of global aggregation rounds as in WFALD, and power control for model transmission is based on  scaled channel inversion as in \cite{zhu2019broadband}.
All the results are averaged over 100 experiments.

\begin{figure}[t]
\centering
\includegraphics[width=6cm]{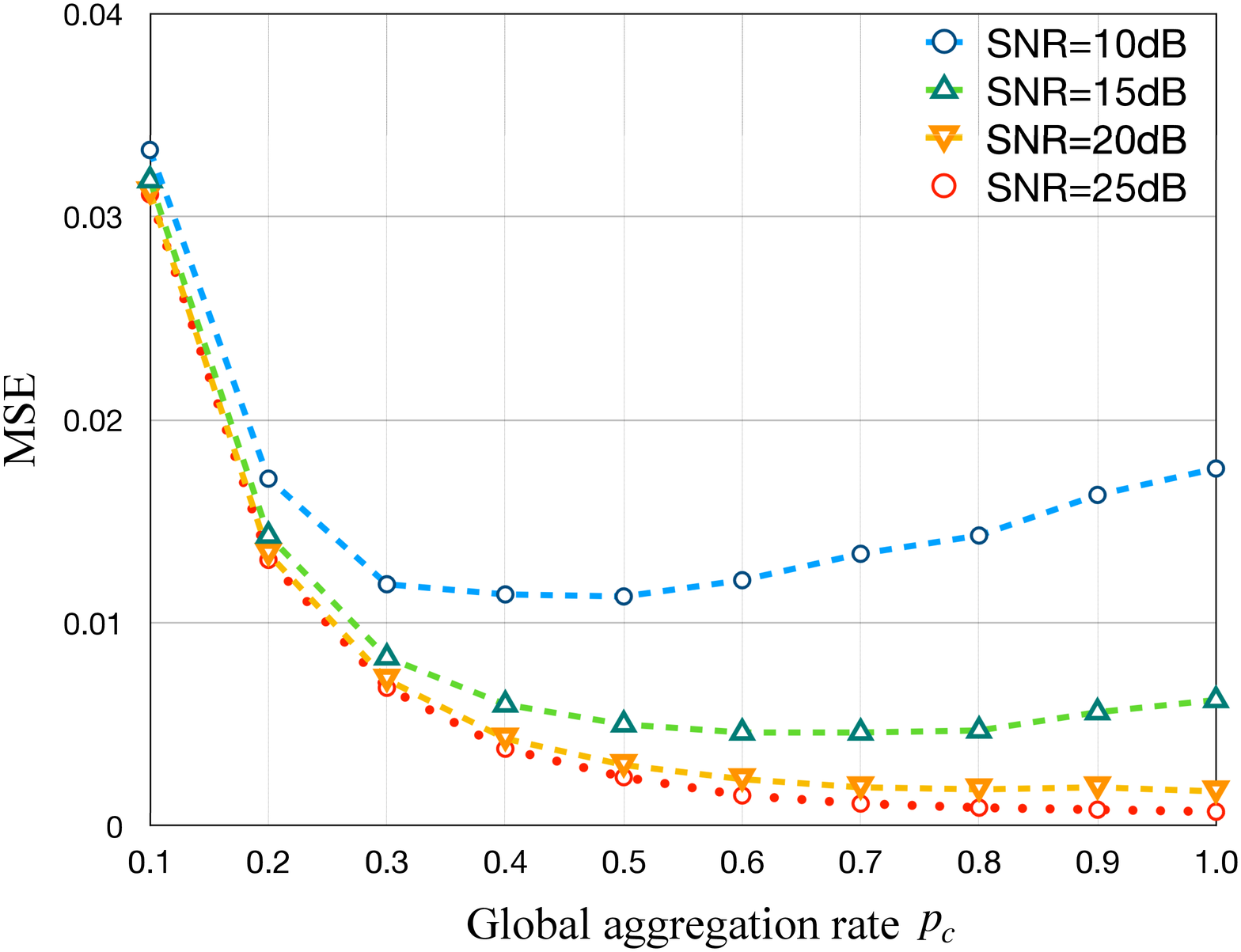}
\vspace{-3mm}
\caption{MSE (\ref{eq:MSE}) as a function of the global aggregation rate $p_c$ for different SNR levels.}
\vspace{-3mm}
\label{Fig: Pc}
\end{figure}

We first investigate the impact of the global aggregation rate $p_c$ on the performance of WFALD by showing in Fig. \ref{Fig: Pc} the MSE (\ref{eq:MSE}) as a function of $p_c$ for different values of the signal-to-noise ratio 
 defined as $\text{SNR} = P / (dN_0)$.    Confirming the analysis in Sec. III, when the SNR is large, it is observed that there is no downside in increasing the frequency $p_c$ of the global aggregation steps, as is the case also in the noiseless setting  studied in \cite{plassier2022federated}. In contrast, as also anticipated by our analysis in the previous section, the additional noise introduced by the channel at low SNRs causes the optimal value of $p_c$ to be generally smaller than 1.

\begin{figure}[t]
\centering
\includegraphics[width=6cm]{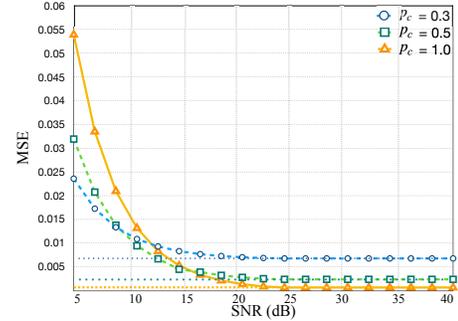}
\vspace{-3mm}
\caption{MSE (\ref{eq:MSE}) as a function of the SNR levels for different global aggregation rate.}
\vspace{-4mm}
\label{Fig: SNR}
\end{figure}

\begin{figure}[t]
\centering
\includegraphics[width=6cm]{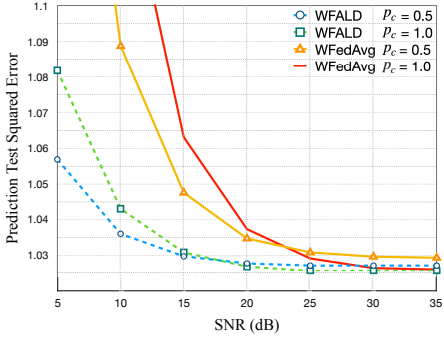}
\vspace{-3mm}
\caption{Prediction test squared error, averaged over all users,  for the ensemble predictor produced by WFALD and for conventional WFedAvg \cite{zhu2019broadband}.}
\vspace{-6mm}
\label{Fig: Comp}
\end{figure}

We study the impact of SNR on the performance of WFALD in Fig. \ref{Fig: SNR}. For the considered global aggregation rates, decreasing SNR causes no harm to the performance until around $25$ dB, with further decreases entailing a performance degradation to the excess noise in the WFALD update. In addition, confirming the analysis in the previous section, we observe that in low-SNR regime, one should carefully choose the aggregation rate $p_c$ to attain an optimal performance. 

Finally, in Fig. \ref{Fig: Comp} we plot the test squared error for the prediction of the variable $v$ obtained with the ensemble predictor provided by WFALD and with the conventional predictor produced by WFedAvg \cite{zhu2019broadband}.  WFLAD demonstrates superior performance as compared to WFedAvg in the considered SNR regime, attaining important gains in the low-SNR regime due to its capacity to leverage channel noise via channel-driven sampling.

\section{Conclusion}
In this paper, we have proposed a novel protocol for the efficient implementation of federated averaging Langevin dynamics (FALD) in wireless systems. As future work, it would be interesting to consider the impact of  data heterogeneity as in \cite{plassier2022federated} and to study larger-scale applications.

\appendix

To prove Theorem 1, the 2-Wasserstein distance is first upper bounded, following \cite[Theorem 1]{dalalyan2017further}, via the  expectation of the 2-norm between the sample generated by Langevin diffusion process  (LDP)  and the sample obtained from WFALD. Then, following similar steps as in proof of \cite[Proposition 1]{WFLMC}, we track separately the contributions of quantization error, stochastic gradient error, and client drift. The analysis of the latter requires a new result as compared to \cite[Proposition 1]{WFLMC}, which is summarized by the following lemma and its proof is similar to \cite[Lemma 10]{plassier2022federated}.  
\begin{lemma}[Upper Bound on the Client Drift]
 \emph{For any iteration $s$, the client drift, defined as 
$V_c^{[s]} =\|  \nabla f(\tilde{\bm \theta}^{[s]}) -  \sum_{k=1}^{K} \hat{\nabla} f_k(\bm{\theta}_k^{[s]}) \|^2$, is upper bounded by  	
 \begin{align}\label{eq: upp client drift}	
 \mathbb{E} [V_c^{[s]}]\leq K^2 L^2 \mathbb{E}[V_{\bm \theta}^{[s]}] +K\sum_{k=1}^{K} \sigma_k^2 ,
 \end{align}
 where we define the local model divergence as $V_{\bm \theta}^{[s]} = \frac{1}{K} \sum_{k=1}^K\|\bm{\theta}_k^{[s]} - \tilde{\bm \theta}^{[s]}\|^2 $. The latter is upper bounded  as 
 \begin{equation}
 \begin{aligned}\label{eq: result ev}
 \mathbb{E}[V_{\bm \theta}^{[s]}] \leq \frac{2(1 - p_c)}{p_c} &\Big[\frac{(2+p_c) \eta^2}{p_c} G^2 +  \frac{\eta^2}{K} \sum_{k=1}^{K} \sigma_k^2   \\ 
 & + \frac{2(K - 1)\eta d}{K} \Big]. 
 \end{aligned}
 \end{equation}
}
\end{lemma}

\end{document}